\documentstyle[12pt]{article}
\newcommand{\tr}[1]{\,{\rm tr}\,#1\,}

\begin{document}
\def\newmathop#1{\mathop{#1}\limits}
\def\ninfty#1{\displaystyle\newmathop{#1}_{N\to\infty}}
\title{
\begin{flushright}
{\small SMI-15-93 \\ November, 1993 }
\end{flushright}
\vspace{2cm}
Regge Regime in QCD \\ and \\ Asymmetric Lattice Gauge Theory}
\author{I.Ya. Aref'eva \thanks{E-MAIL: Arefeva@qft.mian.su}
\\ Steklov Mathematical Institute,\\ Russian Academy of Sciences,\\
Vavilov st.42, GSP-1,117966, \\ Moscow, Russia }
\date{~}
\maketitle
\begin {abstract}

We study the Regge regime of QCD as a special regime of
lattice gauge theory on an asymmetric lattice.
This lattice has a spacing $a_0 $ in the longitudinal direction
and a spacing $a_t $ in the transversal direction.
The limit $\frac{a_{0}}{a_{t}} \to 0$ corresponds to correlation
functions with small longitudinal and large transversal coordinates,
i.e. large $s$ and small $t$.
On this lattice the longitudinal dynamics is described  by the usual
two-dimensional chiral field in finite volume and the transversal
dynamics is emerged through an effective interaction of boundary terms
of the longitudinal dynamics. The effective interaction depends crucially
on the spectrum of the two-dimensional chiral field.
Massless exitations produce an effective 2-dimensional action which
is different from the action recently proposed by H.Verlinde and E.Verlinde.
Massive exitations give rise to an effective action located
on the contour in the longitudinal plane.
\end {abstract}

\newpage

\section{Introduction}
In recent years, efforts have been made to obtain an appropriate scheme to
study QCD in the Regge regime of large energies $\sqrt{s}\to \infty$
and fixed momentum transfers $q (q^{2}=-t)$, $|q|\sim $ $1Gev$, i.e.
$|q|>\Lambda _{QCD}\sim 100 Mev$. It was recognized  \cite {Nach}
that in this regime neither QCD perturbation
theory applies - since $|t|$ is too small -, nor can the usual
lattice gauge theory
approach give numerical answers directly - since $s$ is too large.

In spite of the fact that strictly speaking
studies  of QCD in the Regge regime requires a non-perturbative treating
it is rather instructive to analyze this region within the usual perturbation
expansion. One of the striking results of intensive studies
of this region \cite {LipR,ChWu} is the Lipatov suggestion that
the scattering amplitudes
can be obtained from a two-dimensional field theory \cite {Lip}.

The conclusion that the scattering amplitudes
in the Regge regime are related to two-dimensional field
theories \cite {Lip} came
from using the leading logarithmic approximation (LLA) together with an
incorporation of unitarity condition, i.e. it is in some sense also
non-perturbative. It restores the Froissart boundary:
$\sigma _{tot}<c\ln ^{2}s$,
which is violated in LLA.  The  Froissart boundary can also be restored
in the eikonal approximation. Some years ago, 't Hooft derived the eikonal
approximation for the two-particle scattering amplitude arising from their
gravitation interaction  \cite {'t H} (see also  \cite {Amati}).
Then this problem has been reexamined by
H.Verlinde and E.Verlinde  \cite {VVgr}.  The eikonal approximation
for the two-particle scattering amplitude in QED was studied by Jackiw et al
 \cite {Jackiw}. They have shown that high energy eikonal electrodynamics
can be given an
action formulation where the action is localized on some contour.
The picture of diffraction scattering related to eikonal approximation
was applied in the Regge regime in QCD  by Nachtmann  \cite {Nach}.
According this picture the scattering process between two hadrons can be
represented as a collection of separate collisions between the individual
quarks, and, moreover,
s-dependence of the amplitudes of quark anti-quark scattering
$\Psi _{i}(p_{1})$ + $\Psi _{j}(p_{2})$ $ \to$ $ \Psi _{k}(p_{3}) +
\Psi _{l}(p_{4})$
can be treated in the eikonal approximation
\begin {equation} 
                                                          \label {qsa}
<\Psi _{i}(p_{1}),\Psi _{j}(p_{2})|S|\Psi _{k}(p_{3}),\Psi _{l}
(p_{4})>_{diffr}=
i\delta _{i_{1},i_{3}}
\delta _{i_{2},i_{4}}
\bar{u}(p_{3})\gamma ^{\mu} u(p_{1})
\bar{u}(p_{4})\gamma ^{\mu} u(p_{2})J(q^{2})
\end   {equation} 
and t-dependence is governed by gluon string operators
\begin {equation} 
                                                          \label {csf}
J(q^{2})=-\int d^{2}z e^{iqz}<\tr [{\cal V}_{-}(\infty, 0,
z)-1]\tr[{\cal V}_{+}(0,\infty ,0 )-1]>_{A}
\end   {equation} 
\begin {equation} 
                                                          \label {sf}
{\cal V}_{\pm}(y_{+},y_{-},z)=
P\exp \int _{-\infty}^{y^{\pm}} dy'^{\pm} A_{\pm}(y^{+},y^{-},z),
\end   {equation} 
Here $\pm \equiv \frac{1}{\sqrt{2}}(0\pm 1)$ are used for light-cone
components, $i_{j}$ are colours indices, $<>_{A}$ means an average over
gauge fields. Renormalizability of (\ref {sf}) was proved in  \cite {Acr}.

As it was noted in  \cite {Nach} $J(q^{2})$ should be calculated
non-perturbatively. A simple model to calculate $J(q^{2})$ within
the semi-classical approximation has been proposed by H.Verlinde
and E.Verlinde  \cite {VV}. They  formulated a simple model
in which the two-dimensional nature of the interaction is manifest.
  In our previous paper the
lattice version of the
truncated model  \cite {VV} was investigated  \cite {Ar}.

We would like to stress that one cannot use the standard continuous
perturbative gauge theory in the Regge regime. Note also that the standard
Wilson  lattice calculations  such as strong
coupling expansion or Monte-Carlo calculations in the Regge region also
are not applicable since $s$ is too large. These two at first sight
incompatible circumstances can be made compatible if we start
from an asymmetric lattice.

In this paper we propose to study the Regge  regime as a special regime of
lattice gauge theory on the asymmetric lattice. This lattice has a spacing
$a_0 $ in the longitudinal direction and a spacing
$a_t $ in the transversal direction. Putting $\frac{a_{0}}{a_{t}} \to 0$
we get a possibility to study correlation functions with small
longitudinal and large transversal coordinates, i.e. large longitudinal
and small transversal  components of momentum in the c.m. frame.
We will see that the longitudinal dynamics on this lattice  is described  by
the usual two-dimensional chiral field on the corresponding
two-dimensional lattice. To remove the
lattice in the two-dimensional chiral model we have to perform a
renormalization. It is known that due to dimensional transmutation this
renormalization produces a mass. For example, for the  non-linear
$O(N)$ $\sigma$-model in the large N limit one  gets N massive free
fields with mass $m=Me^{-4\pi /\gamma}$  \cite {Br,An}.

If this two-dimensional chiral model contains massless excitations than
the effective interaction is reduced to an interaction of three discrete
point (two degrees of freedom after taking into account gauge invariance)
and we get a  reduction of the initial 4-dimensional model to
a two-dimensional model containing two fields \footnote{This degrees of
freedom counting agrees with our previous estimations performed within an
approximation scheme which takes into account one-dimensional excitations
 \cite {Ar}.}.
This degrees of
freedom counting disagrees with the H.Verlinde and E.Verlinde  \cite {VV}
counting of
fields and as a result we get an effective action which is different from the
H.Verlinde and E.Verlinde (VV) effective action. If the two-dimensional
chiral model contains massive excitations, as it can be
expected from the dimensional transmutation, then an effective interaction of
boundary terms  cannot be reduced to an interaction of finite
number of degrees of
freedom since there are extra terms in the effective action. These extra terms
are located on some contour in the longitudinal plane
 and a two dimensional reduction is rather
problematic. However under some  special assumptions
this reduction is still possible.

So, our strategy will be the following. We will start from  asymmetric
lattice gauge theory which we will study in the limit
$\frac{a_{0}}{a_{t}} \to 0$. For the resulting theory being the chiral
field model we use the non-perturbative
knowledge of the behaviour of the theory in the continuous limit $a_{0} \to 0$
after suitable renormalization.
Then we will perform the Wick rotation and will approximate boundary
effects in the finite volume D=2 chiral field model by  boundary effects
of D=2  Klein-Gordon theory.
We also incorporate the picture of diffraction scattering for
amplitudes of quark anti-quark scattering (\ref {qsa}).
The calculation  of the correlation
functions of string operators  for  two longitudinal Wilson lines (\ref {csf})
on our asymmetric lattice will be reduced to  examination of
boundary effects in finite volume two-dimensional Klein-Gordon theory.
We will conclude with the
presentation of the effective action for the transversal theory and
we will compare it with the VV effective action.  As it has been mentioned
above our effective action is  different from the VV effective action.
This difference comes from the fact
that for solutions of the wave  equation there are connections
between values of the field on characteristics. Only values of the field
on two characteristics can be treated as independent variables.

In conclusion  we discuss a connection between our asymmetric
lattice gauge theory and Bardeen, Pearson and Rabinovici \cite {Bardeen}
and  Klebanov and Susskind  \cite {Sussk} light-cone lattice gauge theory.

\section{Asymmetric Lattice Gauge Theory}
Since we intend to study the theory in the region of  small longitudinal
coordinates and large transversal
coordinates and we want to take into account fluctuations in different
direction with an approximately equal precision it is relevant to assume
that the lattice spacing in the longitudinal and transversal
directions are different so that there are
equal number of points in the different directions (see fig 1).
If the number of points
in different directions is the same, i.e. $n_{0}=n_{1}=...n_{D-1}$
then we get a theory in an asymmetric space-time volume
$L_{0}L_{1}...L_{d-1}$, $L_{i}$ is a typical size in the i-direction,
with lattice spacing $a_{0},a_{1},...a_{D-1},$ such that
$L_{0}/L_{1}=a_{0}/a_{1}$,
..., $L_{0}/L_{D}=a_{0}/a_{D}$, .

A general form of the lattice
action on an asymmetric lattice with lattice spacing $a_{0},a_{1},...,a_{D-1}$
in 0,1,...D-1-directions has the form
\begin {equation}
                                                          \label {aa}
S=\frac{1}{4g^{2}}a_{0}a_{1}...a_{D-1}\sum _{x,\mu, \nu  }
\frac{1}{(a_{\mu}a_{\nu})^{2}}\tr (U(\Box _{\mu, \nu })-1),
\end   {equation}
Here $x$ are  points of the 4-dimensional lattice,
$\Box _{\mu , \nu }$  is a single plaquette
attached to the links $(x,x+\mu)$ and $(x, x+\nu)$, $U(\Box _{\mu , \nu })=
U_{x,\mu}U_{x+\mu,\nu}U^{+}_{x+\nu,\mu}U^{+}_{x,\nu}$ and link variables
$U_{x,\mu}$ are associated with the link between the lattice sites $x$ and
$x+\hat {\mu}$. $U_{x,\mu}$ belongs to a representation of
the gauge group $SU(N)$.

Since we are interested in the case when $L_{0}=L_{1}$, $L_{2}=L_{3}$ and
$L_{1}/L_{3}\to 0$ we put in (\ref {aa})  $a_{0}=a_{1}$, $a_{2}=a_{3}$
and $a_{0}=\lambda a_{3}$, so we get
\begin {equation}
                                                          \label {al}
S=\frac{1}{4\lambda ^{2}g^{2}}\sum _{x}\sum _{\alpha,\beta }
\tr (U(\Box _{\alpha,\beta })-1) +
\frac{1}{4g^{2}}\sum _{x}\sum _{\alpha, i }
\tr (U(\Box _{\alpha, i })-1)
\end   {equation}
$$+\frac{\lambda ^{2}}{4g^{2}}\sum _{x}\sum _{i,j }
\tr (U(\Box _{i,j})-1).$$
$\alpha,\beta $ are unit
vectors in the longitudinal direction and $i,j$ are unit vectors in
the transversal direction. We also denote the
points of 4-dimensional lattice as $x=(y,z)$, where $y$ and $z$
are the points of two two-dimensional lattices, say, y-lattice (longitudinal)
and z-lattice (transversal).
Performing the $\lambda \to 0$ limit in the lattice action (\ref {aa})
we get
\begin {equation} 
                                                          \label {a}
S^{tr}_{l,l}=\frac{1}{4g^{2}}\sum _{x}\sum _{\alpha, i }
\tr (U(\Box _{\alpha, i }) -1),
\end   {equation} 
with $U_{x,\alpha}$ being a subject of the relation
\begin {equation} 
                                                          \label {zc}
U(\Box _{\alpha,\beta })=1,
\end   {equation} 
i.e. $U _{x,\alpha }$ is a zero-curvature lattice gauge field
\begin {equation} 
                                                          \label {z}
U _{x,\alpha }=V _{x}V ^{+}_{x+\alpha }.
\end   {equation} 
Substituting (\ref {z}) in (\ref {a}) we get
\begin {equation} 
                                                          \label {1}
S^{tr}_{l,l}=\frac{1}{4g^{2}}\sum _{x}\sum _{\alpha, i }
\tr (V_{x+\alpha}^{+}U_{x+\alpha,i}V_{x+\alpha +i}
V^{+}_{x+i}U^{+}_{x, i }V_{x}-1).
\end   {equation} 
or
\begin {equation} 
                                                          \label {A}
S^{tr}_{l,l}=\frac{1}{4g^{2}}\sum _{x}\sum _{\alpha, i }
\tr (\tilde {U}_{x+\alpha,i} \tilde {U}_{x,i}-1)
\end   {equation} 
where
\begin {equation} 
                                                          \label {V}
\tilde {U}_{x}=V^{+}_{x+i}U^{+}_{x, i }V_{x}.
\end   {equation} 
The action (\ref {A}) is ultra-local in the transverse z-direction
and is the lattice chiral field action in the longitudinal y-direction.
In the formal
continuous limit $a_{0}\to 0$ we get
\begin {equation} 
                                                          \label {cl}
S^{tr}_{c,l}=\frac{1}{4g^{2}}\int dy^{0}dy^{1}\sum _{y, i }
\tr[\partial _{\alpha}\tilde {U}_{z,i}(y)\partial _{\alpha}
\tilde {U}^{+}_{z,i}(y)]
\end   {equation} 
with summation over the repeating indexes $\alpha =0,1$ and
$\tilde {U}_{z,i}(y)=V^{+}_{z+i}(y)U^{+}_{z, i }(y)V_{z}(y)$.
To get its pure continuous version one has
also to make in (\ref {cl}) and (\ref {V}) a formal limit $a_{t}\to 0$
under assumption $U_{x,i}=exp(a_{t}A_{i})$
\begin {equation} 
                                                          \label {cc}
S^{tr}_{c,c}=\frac{1}{4g^{2}}\int dy^{0}dy^{1}d^{2}z_{i} \sum _{z,i }
\tr[\partial _{\alpha}\tilde {A}\partial _{\alpha}\tilde {A}]
\end   {equation} 
where
\begin {equation} 
                                                          \label {VV}
\tilde {A}_{i}=V^{+}D_{i}V,~~ D_{i}=\partial _{i}+A_{i},
\end   {equation} 
 \begin{figure}
\setlength{\unitlength}{0.5cm}
\begin{center}
\begin{picture}(5,5)(-1,-2)
 \multiput(0.0,0.0)(0.5,0.0){5}{\line(0,1){4.0}}
 \multiput(0.0,0.0)(0.0,1.0){5}{\line(1,0){2.0}}
 \put(-1.0,1.0){$a_{2}$}
\put(0.0,-1.0){$a_{1}$}
\put(2.0,-1.0){$L_{1}$}
\put(-1.0,4.0){$L_{2}$}
 \put(-2.0,-2.5){$n_{1}=n_{2},~~\frac{L_{1}}{L_{2}}=\frac{a_{1}}{a_{2}}$}
\end{picture}
\end{center}
\caption{Asymmetric lattice }\label{f1}
\end{figure}
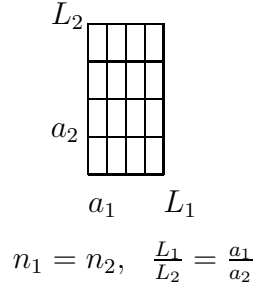

Therefore in the formal limit $a_{t} \to 0 $ the action (\ref {cl})
reproduces the  Verlinde and
Verlinde truncated action. However, as we will see in the next section,
non-perturbatively there is an essential
difference in the behaviour of two theories (\ref {cc}) and (\ref {cl}).
Note that the action (\ref {al}) was considered in  \cite {Ar} as
a discrete version of the
 Verlinde and Verlinde rescaling  action.

It is also instructive to compare the action (\ref {cl}) with the
Bardeen, Pearson and Rabinovici   lattice action \cite {Bardeen} for
light-cone gauge theory. They also considered $a_{0}\to 0$ limit of the
Willson lattice theory but in addition they assumed that the gauge field
on the longitudinal links fluctuate around the unit element of the group, i.e.
$U_{x,\alpha}=\exp (a_{0}A_{\alpha})$. Their action has the form
\begin {equation} 
                                                          \label {BA}
S_{c,l}=\frac{1}{g^{2}}\int dy^{0}dy^{1}[
\sum _{z}\sum _{\alpha, \beta }\frac{a_{t}^{2}}{2}
\tr (F_{\alpha, \beta }F^{\alpha, \beta }) +
\end   {equation} 
$$\sum _{z}\sum _{\alpha, i }
\tr (D_{\alpha}U_{z,i}(y)D^{\alpha}U^{+}_{z,i}(y))
+ \sum _{z}\sum _{i,j }\frac{1}{a_{t}^{2}}\tr (U(\Box_{i,j})-1)].
$$
Examining the above formulae (\ref {cl}) and (\ref {BA})  shows that the action
(\ref {cl}) with the zero-curvature condition (\ref {z}) can be extracted from
(\ref {BA}) in the limit of large transversal lattice spacing $a_{t}$.

The lattice version of (\ref {sf}) for the zero curvature  longitudinal
gauge fields is simply reduced to the boundary values of the field $V_{z}(y)$
\begin {equation} 
                                                          \label {lsf}
{\cal V}_{+}(L,0,z) =V_{z}(-L,0)V^{+}_{z}(L,0),~~
{\cal V}_{-}(0,L,z) =V_{z}(0,-L)V^{+}_{z}(0,L).
\end   {equation} 
and the expectation value of these string operators is given by
\begin {equation} 
                                                          \label {EA}
<{\cal V}_{+}(L,0,0){\cal V}_{-}(0,L,z)>=
\end   {equation} 
$$\int V_{0}(-L,0)V^{+}_{0}(L,0)
V_{z}(0,-L)V^{+}_{z}(0,L)\exp S^{tr}(V,U) dV dU$$
with $S^{tr}(V,U)$ as in (\ref {A}). In the continuum limit in the
longitudinal direction we use $S^{tr}$ given by (\ref {cl}).
Our goal consists in calculation the functional integral (\ref {EA})
over gauge fields U to get an effective action describing an interaction
of fields $V_{z}(0,\pm L)$ $V_{z}(\pm L,0)$.

\section{Boundary Effects for D=2  Klein-Gordon Action}
\subsection{Boundary Effects  for D=2 massless case}
Let us assume that the excitation spectrum of the theory (\ref {cl})
contains massless
particles. This means that in this case we can forget that our fields
variables are restricted by the unitarity condition and we can deal with
the action  for the wave equation
\begin {equation} 
                                                          \label {M}
S=\int dy^{0}dy^{1}
\tr[\partial _{\alpha}\tilde {M}(y)\partial _{\alpha}\tilde {M}^{+}(y)],
\end   {equation} 
where $\tilde {M}_{z,i}(y)=V^{+}_{z}(y)M_{z,i}(y)V_{z+i}(y)$ and we drop
for the  moment indexes $z$ and $i$. $\tilde {M}(y)$ in (\ref {M}) is
an arbitrary complex matrix. By the usual Feynman integral
prescription  one can compute the transition function from one given
configuration $M_{a}$  at the initial time to another configuration
$M_{b}$ at time $y^{0}$.
Since we are interested in special correlation functions related with
the values of $M_{z,i}(y^{+},y^{-})$  at  $y^{\pm}=\pm L,$ $ y^{\mp}=0$
respectively, it is more convenient to consider the transition function
in the light-cone variables.

Let us consider for a moment the transition function for one component
massless field in the
light-cone variables
\begin {equation} 
                                                          \label {K}
{\cal K}(\phi _{1}(y^{+}),\phi _{2}(y^{-}),\phi _{3}(y^{+}),\phi _{4}(y^{-}))=
\int \exp \{ i S(\phi ,L)\}\cdot \prod
d\phi (y^{+}y^{-}),
\end   {equation} 
\begin {equation} 
                                                          \label {FS}
S(\phi ,L)=\int _{-L}^{L}dy^{+}\int _{-L}^{L}dy^{-}\partial _{-}\phi
(y^{+}y^{-})
\partial _{+}\phi (y^{+}y^{-}),
\end   {equation} 
where the following boundary conditions are assumed (see fig.2)
\begin{figure}
\setlength{\unitlength}{0.5cm}
\begin{center}
\begin{picture}(10,10)(-5,-5)
  \put(-5.0,0.0){\vector(1,0){12.0}}
 \put(0.0,-5.0){\vector(0,1){12.0} }
\put(-2.5,-2.5){\vector(1,1){6.0} }
\put(2.5,-2.5){\vector(-1,1){6.0} }
        \multiput(-5.0,0.0)(5.0,5.0){2}{\line(1,-1){5.0}}
 \put(-5.0,0.0){\line(1,1){5.0}}
 \put(0.0,-5.0){\line(1,1){5.0}}
 \put(0.5,7.0){\mbox {$y^{0}$}}
 \put(7.5,0.0){\mbox {$y^{1}$}}
 \put(3.5,3.0){\mbox {$y^{+}$}}
 \put(-4.5,3.0){\mbox {$y^{-}$}}
  \put(1.2,4.0){\mbox {$\phi _{4}(y^{-})$}}
  \put(-3.5,4.0){\mbox {$\phi _{3}(y^{+})$}}
\put(0.5,5.0){\mbox {{\bf IV}}}
\put(-5.5,-0.8){\mbox {{\bf I}}}
\put(5.2,-0.8){\mbox {{\bf III}}}
\put(0.5,-5.5){\mbox {{\bf II}}}
\put(-2.0,2.2){\mbox {{\bf 3}}}
\put(1.5,2.2){\mbox {{\bf 4}}}
\put(1.5,-2.7){\mbox {{\bf 1}}}
\put(-1.8,-2.7){\mbox {{\bf 2}}}
\put(1.5,-4.0){\mbox {$\phi _{1}(y^{+})$}}
 \put(-4.0,-4.0){\mbox {$\phi _{2}(y^{-})$}}
\put(-5.2,-2.0){\mbox {$\Gamma _{I,II}$}}
\put(4.0,-2.0){\mbox {$\Gamma _{II,III}$}}
\put(2.5,-2.5){\circle*{0.2}}
\put(2.5,2.5){\circle*{0.2}}
\put(-2.5,-2.5){\circle*{0.2}}
\put(-2.5,2.5){\circle*{0.2}}
\put(5.0,0.0){\circle*{0.2}}
\put(-5.0,0.0){\circle*{0.2}}
\put(0.0,5.0){\circle*{0.2}}
\put(0.0,-5.0){\circle*{0.2}}
\end{picture}
 \end{center}
\caption{ Boundary conditions}\label{f2}
\end{figure}
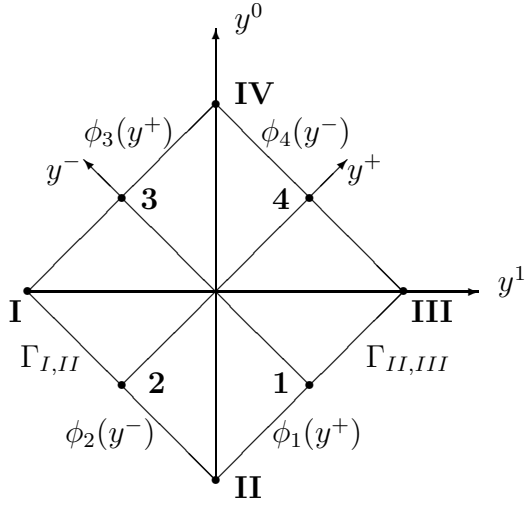

\begin {equation} 
                                                          \label {BC}
\phi (y^{+},y^{-})|_{y^{-}=-L}=\phi _{1}(y^{+}),~
{}~\phi (y^{+},y^{-})|_{y^{+}=-L}=\phi _{2}(y^{-}),
\end   {equation} 
\begin {equation} 
                                                          \label {BC'}
\phi (y^{+},y^{-})|_{y^{-}=L}=\phi _{3}(y^{+}),~~
\phi (y^{+},y^{-})|_{y^{+}=L}=\phi _{4}(y^{-})
\end   {equation} 

The boundary of the space-time region in (\ref {FS}) consists on four
characteristics.
The classical solution of the wave equation can be specified by functions
on two characteristics  (the Goursat boundary
problem), say $\phi _{1}$ and $\phi _{2}$, having the same value on
a common point  $\phi _{1}(-L)=\phi _{2}(-L)$
\begin {equation} 
                                                          \label {S}
\phi _{cl}(y^{+},y^{-})=\phi _{1}(y^{+})+\phi _{2}(y^{-}) -\phi _{1}(-L)
\end   {equation} 

Other two functions being  values of the solution of the wave equation
on others two
characteristics can be written in terms of $\phi _{1}$ and $\phi _{2}$ as
\begin {equation} 
                                                          \label {3}
\phi ^{cl}_{3}(y^{+})\equiv\phi _{cl}(y^{+},y^{-})|_{y^{-}=L}=
\phi _{1}(y^{+})+\phi _{2}(L) -\phi _{2}(-L),
\end   {equation} 
\begin {equation} 
                                                          \label {4}
\phi ^{cl}_{4}(y^{-})\equiv\phi _{cl}(y^{+},y^{-})|_{y^{+}=L}=
\phi _{1}(L)+\phi _{2}(y^{-}) -\phi _{1}(-L).
\end   {equation} 
The transition function (\ref {K})
is proportional to $\delta$-functions guaranteeing that $\phi _{3}$ is equal to
$\phi ^{cl}_{3}$  and $\phi _{4}$ is equal to $\phi ^{cl}_{4}$, i.e.
\begin {equation} 
                                                          \label {D}
{\cal K}(\phi _{1}(y^{+}),\phi _{2}(y^{-}),\phi _{3}(y^{+}),
\phi _{4}(y^{-}))=
\prod _{-L\leq y^{+}\leq L}\delta (\phi _{3}(y^{+})-\phi ^{cl}_{3}(y^{+}))
\cdot
\end   {equation} 
$$\prod _{-L\leq y^{-}\leq L}\delta (\phi _{4}(y^{-})-\phi
^{cl}_{4}(y^{-})) \exp \{S(\phi _{cl},L)\} $$
where  $S(\phi _{cl},L)$ is given by the classical action being calculated
on the classical solution (\ref {S}). $S(\phi _{cl},L)$ has the form
\begin {equation}
                                                       \label {CA}
S(\phi _{cl},L)=(\phi _{1}(L)-\phi _{1}(-L))\cdot(\phi _{2}(L)-\phi _{1}(-L))
\end   {equation} 
The representation (\ref {D})  can be directly verifyed from (\ref {K}) by
using the symmetric approximation of the functional integral
in the LHS of (\ref {K}).

It is turn out that the classical action depends only on the values of the
field $\phi$ on tree points, namely, on points I, II and III on fig.2.
Of course, the action  (\ref {FS}) on the classical solution can be
written in terms of values of
the classical solution on four points, say points 1,2,3 and 4 on fig.2
as it was done in  \cite {VV},
\begin {equation} 
                                                          \label {CAVV}
S(\phi _{cl},L)=(\phi _{cl}(L,0)-\phi _{cl}(-L,0))\cdot
(\phi _{cl}(0,L)-\phi _{cl}(0,-L)).
\end   {equation} 
However for the classical solutions there are two constraints which make
the differences
$(\phi _{cl}(L,y^{-})$-$\phi _{cl}(-L,y^{-}))$ and $(\phi _{cl}(y^{+},L)-
\phi _{cl}(y^{+},-L))$ not depending on $y^{-}$ and $y^{+}$, respectively.
We can put $y^{-}=-L$ and $y^{+}=-L$, making this action depending explicitly
on the field $\phi$ on tree points, but not on four points as in
\cite {VV}.

Specifying the values of the
field $\tilde {M} _{z,i}(y^{+}y^{-})$ on these tree points as
$\tilde {M} ^{A}_{z,i}$,
$A=I,II,III$ we get an effective action
\begin {equation} 
                                                          \label {EAM}
S_{eff}=\sum _{z,i}\tr (\tilde {M}^{I} _{z,i}-\tilde {M}^{II} _{z,i})
(\tilde {M}^{III} _{z,i}-\tilde {M}^{II} _{z,i})
\end   {equation} 
and therefore
\begin {equation} 
                                                          \label {LEA}
S_{eff}=\sum _{z,i}\tr (V_{z}^{+}(I)U_{z,i}(I)V_{z+i}(I)
-V_{z}^{+}(II)U_{z,i}(II)V_{z+i}(II))
\end   {equation} 
$$
\cdot (V_{z}^{+}(III)U_{z,i}(III)V_{z+i}(III)
-V_{z}^{+}(II)U_{z,i}(II)V_{z+i}(II)).$$
In contrast to  \cite {VV} the effective action (\ref {EAM}) depends
only on three two-dimensional chiral fields $V^{A} _{z}$, $A=I,II,III$.

In the continuous limit $a_{t}\to 0$ the effective action (\ref {EAM})
yields
\begin {equation} 
                                                          \label {EZM}
S_{eff}=\int d^{2}z \tr (V^{+}(I,z)D^{I}_{i}V(I,z)-
V^{+}(II,z)D^{II}_{i}V(II,z))
\end   {equation} 
$$\cdot
(V^{+}(III,z)D^{III}_{i}V(III,z)-V^{+}(II,z)D^{II}_{i}V(II,z)).$$

\subsection{Boundary Effects  for D=2 massive case}
Unfortunately we cannot restrict ourselves only by massless excitations.
The spectrum of the chiral model is more complicated. If we knew the exact
behaviour of the chiral  model, we could rewrite the action in terms of
linear (unrestricted) variables without approximations. Now we assume
that non-perturbative spectrum contains massive excitations.
A more rigorous argument for the above statement comes from the knowledge
of the behaviour of the non-linear O(N) $\sigma$-model in two dimensions.
This model in the leading large N order can be described by N-component
massive free field with mass $m=Me ^{-4\pi /g^{2}}$.
This gives us a reason to consider boundary effects for free massive theory
\footnote{Since we consider a theory in finite space-time it is
natural to expect that the mass renormalization will depend on the volume.
The question was analyzed by Luscher some years ago  \cite {Luscher}. It was
found that finite volume effects on the mass renormalization
are related to forward scattering amplitudes of the infinite volume theory.}
\begin {equation} 
                                                          \label {MA}
S=\int dy^{0}dy^{1}
\tr[\partial _{\alpha}\tilde {M}(y)\partial _{\alpha}\tilde {M}^{+}(y)+
m^{2}\tilde {M}(y)\tilde {M}^{+}(y)],
\end   {equation} 
as an approximation to boundary effects  for the model (\ref {cl}).

For the massive
transition function given by the functional integral
\begin {equation} 
                                                          \label {KM}
{\cal K}_{m}(\phi _{1}(y^{+}),\phi _{2}(y^{-}),
\phi _{3}(y^{+}),\phi _{4}(y^{-}))=\int \exp \{ S_{m}(\phi , L)\}
  \prod d\phi (y^{+}
y^{-})
\end   {equation} 
with an action
\begin {equation} 
                                                          \label {MA''}
S_{m}(\phi , L)= \int _{-L}^{L}dy^{+}\int _{-L}^{L}dy^{-}
(\partial _{-}\phi (y^{+}y^{-})
\partial _{+}\phi (y^{+}y^{-}) + m^{2}\phi ^{2}(y^{+}y^{-}))
\end   {equation} 
and with boundary conditions (\ref {BC}) and (\ref {BC'})
also takes place the representation (\ref {D}).

The value of the massive action on classical solutions can be written in
terms of  values of the classical field on four points I,II,III and IV in
fig.2
\begin {equation} 
                                                          \label {MA'}
S_{m}(\phi _{cl}, L) = \frac{1}{2}[\phi _{cl}^{2} (L,L)-\phi _{cl}^{2} (-L,L)-
\phi _{cl}^{2} (L,-L)+\phi _{cl}^{2} (-L,-L)]
\end   {equation} 

But there is an essential difference with the massless case. Instead of
the representation of the classical solutions in the form (\ref {S}) we have
\begin {equation} 
                                                          \label {MS}
\phi _{m}^{cl}(y^{+}, y^{-})=\phi _{1}(y^{+})+\phi _{2}(y^{-}) -
\phi _{1}(-L)R(y^{+}, y^{-};-L,-L)
-\end   {equation} 
$$\int _{-L}^{y^{+}} d\xi \phi _{1}(\xi)\partial _{\xi}
R(y^{+}, y^{-};\xi,-L)
-\int _{-L}^{y^{-}} d\eta \phi _{1}(\eta)\partial _{\eta}
R(y^{+}, y^{-};-L,\eta),
$$
where the Riemann function $R$ can be written in terms of the Bessel
function $J_{0}$
\begin {equation} 
                                                          \label {RF}
R(y^{+}, y^{-};\xi,\eta)=J_{0}(\sqrt{4m^{2}(\xi -y^{+})(\eta -y^{-})}).
\end   {equation} 
Note that for $m=0$ one has $R=1$ and the formula
(\ref {MS}) reproduces the solution (\ref {S}). $\phi _{cl}(L,L)$
depends on the full set
of the initial data of the Goursat boundary problem and after substitution of
$\phi _{cl}(L,L)$ in the action (\ref {MA}) we get
\begin {equation} 
                                                          \label {MAE}
S_{m}(\phi _{cl},L) =\frac{1}{2} \{-
\phi _{1}^{2} (L)-\phi _{2}^{2} (L) +\phi _{1}^{2} (-L)
\end   {equation} 
$$+ [\phi _{1} (L)+\phi _{2} (L)-\phi _{1} (-L)
J_{0}(4mL)-
\int _{-L}^{L} d\xi (\phi _{1}(\xi)+\phi _{2}(\xi))
\partial _{\xi}J_{0}(\sqrt{8m^{2}L(L-\xi )}~ ]^{2}\}
$$
Let us now examine the effective action (\ref {MAE}). We see that this
effective action is located on the contours $\Gamma _{I,II}$ and
$\Gamma _{II,III}$ in the $(y^{+},y^{-})$-plane.

If we assume that the mass $m$ is fixed and $L$ goes to infinity and
$\phi _{1}$ and $\phi _{2}$
vary smooth enough on the interval (-L,L) then we can neglect the terms with
the Bessel function and  we get
\begin {equation} 
                                                          \label {MAA}
S(\phi _{cl}) =
\frac{1}{2}[(\phi _{1} (L)\cdot\phi _{2} (L)+\phi ^{2}_{1} (-L)]
\end   {equation} 
So, as in the massless case the effective action is located on tree points,
but the form of action is not the same. Now the effective action is given by
\begin {equation} 
                                                          \label {EM}
S_{eff}=\int d^{2}z \tr (V^{+}(I,z)D^{I}_{i}V(I,z)\cdot
V^{+}(III,z)D^{III}_{i}V(III,z)\end   {equation}
$$+V^{+}(II,z)D^{II}_{i}V(II,z)\cdot
V^{+}(II,z)D^{II}_{i}V(II,z)).
$$

If we assume that the normalization point $M$ in the mass renormalization
is related with $L$ so that $ML$ is fixed for $L\to \infty$ then
in this case we cannot neglect the terms in the effective action
located on the contours $\Gamma _{I,II}$ and
$\Gamma _{II,III}$.

\section{Expectation Value of String Operators}
\subsection{Abelian case}
It is instructive to start consideration of expectation value of string
operators (\ref {sf})  from the abelian case.
We know that the properties of compact and non-compact QED are different
\cite {Pol}. So it is natural to expect that the expectations value of string
operators (\ref {sf}) will be different in compact and non-compact QED.
Below we will present calculations
which take into account only the usual exitations postponing a consideration of
topologically stable quantized-vortex configurations for
future investigations.

In the standard QED the
expectation value of string operators (\ref {sf})
in the Born approximation, i.e. in the approximation neglecting fermionic
loops, is given by
\begin {equation} 
                                                          \label {ase}
< {\cal V}_{-}(L, 0,
z){\cal V}_{+}(0,L ,0 )>_{QED}=\int e^{-\frac{i}{4g^{2}}\int d^{4}x F_{\mu
\nu}^{2}
+\int d^{4}x A_{\mu}J^{\mu}} \prod \delta (\partial _{\mu} A_{\mu} )dA_{\mu}
\end   {equation} 
with
\begin {equation} 
                                                          \label {ac}
J^{\mu}= \int _{\Gamma _{13}} \delta ^{4}(x-x(\sigma))dx^{\mu}(\sigma)+
 \int _{\Gamma _{24}} \delta ^{4}(x-x(\sigma))dx^{\mu}(\sigma)
\end   {equation} 
where the contours $\Gamma _{13}$ and  $\Gamma _{24}$  are specified as
$x^{+}(\sigma)=\sigma$, $x^{-}(\sigma)=0$, $x^{i}(\sigma)=z^{i}$ and
$x^{+}(\sigma)=0$, $x^{-}(\sigma)=\sigma$, $x^{i}(\sigma)=0$, $-L\leq\sigma
\leq L$, respectively. Therefore
\begin {equation} 
                                                          \label {as1}
< {\cal V}_{-}(L, 0,z){\cal V}_{+}(0,L ,0 )>_{QED}=
\end   {equation} 
$$\exp \{\frac{i g^{2}}{2}\int dk_{+}dk_{-}d^{2}k_{i}
\frac{\sin k_{+}L}{k_{+}}\cdot
\frac{\sin k_{-}L}{k_{-}}\cdot \frac{e^{ik_{i}z^{i}}}{k_{+}k_{-}-k_{i}^{2}}
\}
$$
In the limit $L\to \infty $ two first factors in the integrand produce
$\delta (k_{+})\delta (k_{-})$ and we get
\begin {equation} 
                                                          \label {as2}
< {\cal V}_{-}(\infty , 0,z){\cal V}_{+}(0,\infty ,0 )>_{QED}= \exp \{
-\frac{ig^{2}}{2}\int d^{2}k_{i}\frac{e^{ik_{i}z^{i}}}{k_{i}^{2}}
\}
\end   {equation} 
 Thus $J(q^{2})$ given by (\ref {csf})  for QED can be represented as
\begin {equation} 
                                                          \label {as3}
J_{QED}(q^{2})=-\int d^{2}z e^{iqz}(\exp \{ -\frac{i g^{2}}{2}
\int d^{2}k_{i}\frac{e^{ik_{i}z^{i}}}{k_{i}^{2}} \}-1)
\end   {equation} 
Note that if we introduce  infrared regulating mass $\mu $ then (\ref {as3})
for small $\mu$ reproduces the standard eikonal formula
\cite {'t H,Jackiw}
\begin {equation} 
                                                          \label {ar}
J_{QED}(q^{2}) =\frac{\Gamma (1+ig^{2}/4\pi )}{4\pi \mu ^{2}
\Gamma (-g^{2}/4\pi )}
(\frac{4\mu ^{2}}{q^{2}})^{1+ig^{2}/4\pi }
\end   {equation} 

Let us derive this result starting from the action (\ref {cl}). In this case
$V_{z}(y)= e^{i\phi _{z}(y)}$ and for small $a_t~$ $\tilde M _{z,i}(y)=
a_{t}\partial _{i}\phi (y,z)$. According (\ref {EAM})  and we have
the following effective action
\begin {equation} 
                                                          \label {eaa}
S^{QED}_{eff}=\int d^{2}z [\partial _{i}\phi _{1}(L,z)-
\partial _{i}\phi _{1}(-L,z)][\partial _{i}\phi _{2}(L,z)-
\partial _{i}\phi _{1}(-L,z)]
\end   {equation} 
In terms of the phase factor $\phi$  the string operators ${\cal V}_{\pm}$
have the simple form
\begin {equation} 
                                                          \label {asp}
{\cal V}_{+}(0,L,z)= e^{i(\phi (0,L,z)-\phi (0,-L,z))},~~
{\cal V}_{-}(L,0,z)= e^{i(\phi (L,0,z)-\phi (-L,0,z))}
\end   {equation} 
For $\phi (y^{+},y^{-},z)$ being a solution of the wave equation on
$\pm$ variables
the string operators ${\cal V}_{\pm}(z)$ should be written in terms
of the initial date of the Goursat problem
\begin {equation} 
                                                          \label {asg}
{\cal V}_{+}(0,L,z)= e^{i(\phi _{2}(L,z)-\phi _{1}(-L,z))},~~
{\cal V}_{-}(L,0,z)= e^{i(\phi _{1}(L,z)-\phi  _{1}(-L,z))}
\end   {equation} 
Therefore the correlator (\ref {as1}) for QED is
\begin {equation} 
                                                          \label {acp}
< {\cal V}_{-}(L,0,z){\cal V}_{+}(0,L,0)>_{QED}
=\int  \exp \{ i(\phi _{2}(L,z)-\phi _{1}(-L,z))+\phi _{1}(L,0)-
\end   {equation} 
$$\phi  _{1}(-L,0))+
\int d^{2}z [\partial _{i}\phi _{1}(L,z)-
\partial _{i}\phi _{1}(-L,z)][\partial _{i}\phi _{2}(L,z)-
\partial _{i}\phi _{1}(-L,z)]\}$$
$$ \prod _{z} d\phi _{1}(L,z)d\phi _{2}(L,z)
d\phi _{1}(-L,z).
$$
Performing the change of variables
$\rho _{1}(z)=\phi _{1}(L,z)-\phi  _{1}(-L,z)$,
$\rho _{2}(z)=\phi _{2}(L,z)-\phi  _{1}(-L,z)$ and eliminating  gauge degrees
 of freedom we left with the functional integral over two fields
\begin {equation} 
                                                          \label {acp'}
< {\cal V}_{-}(L,0,z){\cal V}_{+}(0,L,0 )>_{QED}=
\end   {equation} 
$$\int \exp \{ i (\rho _{1}(z)-\rho _{2}(z))+
\int d^{2}z \partial _{i}\rho _{1}(z)\partial _{i}\rho _{2}(z)\}
 \prod _{z} d\rho _{1}(z)d\rho _{2}(z)$$
that gives the same answer as the previous calculation (\ref {as3}).

\subsection{Nonabelian case}
In the non-abelian case the string operators (\ref {lsf}) also should be
written in terms of independent variables (the initial data of the Goursat
problem). In the case when we restrict ourselves by the consideration
only massless excitations we have to take into account the following
relation (see fig.2)
\begin {equation} 
                                                          \label {R}
\tilde {M}(4)-\tilde {M}(2)=\tilde {M}(III)-\tilde {M}(II)
\end   {equation} 
So, V(4) should be find as a solution of the difference equation
\begin {equation} 
                                                          \label {dE}
V_{z}^{+}(4)U_{z,i}(4)V_{z+i}(4)=V_{z}^{+}(2)U_{z,i}(2)V_{z+i}(2)
\end   {equation} 
$$+V_{z}^{+}(III)U_{z,i}(III)V_{z+i}(III)-V_{z}^{+}(II)U_{z,i}(II)V_{z+i}(II)
$$
In the continuum case this equation has the form
\begin {equation} 
                                                          \label {DE}
D^{4}_{i}V(4,z)=V(4,z)(V^{+}(2,z)D^{2}_{i}V(2,z)+
\end   {equation} 
$$V^{+}(III,z)D^{III}_{i}V(III,z)-V^{+}(II,z)D^{II}_{i}V(II,z))
$$
A solution of this equation can be represented as
\begin {equation} 
                                                          \label {SDE}
V(4,z)=V(4,0)P\exp \int _{0}^{z}(A_{i}(4,z')+V^{+}(2,z')D^{2}_{i}V(2,z')
\end   {equation} 
$$
+V^{+}(III,z')D^{III}_{i}V(III,z')-V^{+}(II,z')D^{II}_{i}V(II,z'))dz'
$$
The necessity of taking into account the constraint (\ref {R}) makes the
calculation of the quark-antiquark scattering amplitude more complicated
as compare with calculations performed in  \cite {VV} and we postpone
this calculation for future publications. Note here only that
motivated by the previous consideration of the abelian case we can
assume that in a rough approximation
\begin {equation} 
                                                          \label {RA}
V^{+}(4,z)V(2,z)=V^{+}(III,z)V(II,z) +\mbox {corrections}
\end   {equation} 
In this approximation we can reproduce only one type of diagrams describing
the  quark-antiquark scattering amplitude.
\newpage
\section{Concluding Remarks.}
Our paper was stimulated by the Verlinde and Verlinde paper
\cite {VV} but there are some essential differences between our works.

$\bullet $ We start from the asymmetric lattice action, because
the standard QCD perturbation
theory cannot be applied in the Regge regime since $|t|$ is too small.

$\bullet\bullet $ We have obtained that the longitudinal
dynamics of the truncated asymmetric lattice theory is described by the usual
two-dimensional chiral field. Its
continuous version   depends crucially on the spectrum
of the usual two-dimensional chiral field model in finite volume.
In the framework of the perturbation theory there are only
massless exitations and the corresponding continuum version is described by
the free massless field. E.Verlinde and H.Verlinde started from this massless
field that is correct only perturbatively.

$\bullet\bullet\bullet $ Transversal dynamics arising  from boundary
effects for free massless field and chiral field are different.
These two theories coincide only in the leading order of the standard
perturbation theory. Therefore two approaches give the same results
only within the standard perturbation theory. The lattice version permits to
take into account non-perturbative effects.

$\bullet\bullet\bullet\bullet $ Even if we start from free massless
field as E.Verlinde and H.Verlinde did we have to take into account
constrains on the boundary
classical fields in the Feynman light-cone transition functions.
This reduces the number of independent degrees of freedom and makes more
complicated the calculation of string correlation functions in the
non-abelian case.

The central observation of the paper \cite {VV} is that the leading Regge
regime contribution comes from those gauge field configurations that are flat
in the longitudinal directions. Consideration of asymmetric lattice gauge
theory in the limit $a_{0}/a_{t} \to 0$ confirms this statement.

Our limiting theory for the continuous space-time in the longitudinal
directions (\ref {cl}) does not reproduce the light-cone
lattice action \cite {Bardeen} since we do not assume that
the longitudinal gauge field fluctuate around the unit element of the gauge
group.
As it has been mentioned in Sect.2 there is a connection between the
truncated lattice action (\ref {cl}) and the light-cone transversal
lattice QCD action considered previously by Bardeen, Pearson and
Rabinovici  \cite {Bardeen}. Namely,  (\ref {cl}) is the
large $a_{t}$  version of the light-cone  transversal lattice QCD
\cite {Bardeen}. The study of this theory
for large $a_{t}$ corresponds to the regime of small, but
non-zero $|t|$. Note that just this model due to the recent work of
Klebanov and Susskind  \cite {Sussk} has a rather interesting and
mysterious string interpretation.
Perhaps this is not so surprising since fluctuations of some
special excitations of the two-dimensional chiral model, namely one-dimensional
excitations, make this model in some sense equivalent to $\mbox{QCD}_{2}$ which
has string interpretation according to recent Gross and Taylor
investigations \cite {GT}.
$$~$$
{\bf ACKNOWLEDGMENT}
$$~$$
The author is grateful to A.A.Slavnov and I.V.Volovich for useful
discussions.

\end{document}